\documentclass[conference,letterpaper]{IEEEtran}

\usepackage{cite}
\usepackage{amsmath,amssymb,amsfonts}
\usepackage{algorithmic}
\usepackage{graphicx}
\usepackage{textcomp}
\usepackage{xcolor}
\usepackage{booktabs}
\usepackage{float}
\usepackage[hidelinks]{hyperref}
\usepackage{listings}
\usepackage{color}
\definecolor{lightgray}{rgb}{.9,.9,.9}
\definecolor{darkgray}{rgb}{.4,.4,.4}
\definecolor{purple}{rgb}{0.65, 0.12, 0.82}

\lstdefinelanguage{JavaScript}{
  keywords={typeof, new, true, false, catch, function, return, null, catch, switch, var, if, in, while, do, else, case, break},
  keywordstyle=\color{blue}\bfseries,
  ndkeywords={class, export, boolean, throw, implements, import, this},
  ndkeywordstyle=\color{darkgray}\bfseries,
  identifierstyle=\color{black},
  sensitive=false,
  comment=[l]{//},
  morecomment=[s]{/*}{*/},
  commentstyle=\color{purple}\ttfamily,
  stringstyle=\color{red}\ttfamily,
  morestring=[b]',
  morestring=[b]"
}

\def\BibTeX{{\rm B\kern-.05em{\sc i\kern-.025em b}\kern-.08em
    T\kern-.1667em\lower.7ex\hbox{E}\kern-.125emX}}

\begin{document}

\title{Controlled Update of Software Components using Concurrent Exection of Patched and Unpatched Versions}

\author{\IEEEauthorblockN{Stjepan Groš, Ivan Kovačević, Ivan Dujmić, Matej Petrinović}
\IEEEauthorblockA{\textit{Laboratory for Information Security and Privacy} \\
\textit{Faculty of Electrical Engineering and Computing University of Zagreb} \\
Zagreb, Croatia \\
Contact author mail: stjepan.gros@fer.hr}
}
% TODO: Dodati da sam ja kontakt autor!

\maketitle

\begin{abstract}
Software patching is a common method of removing vulnerabilities in software components to make IT systems more secure. However, there are many cases where software patching is not possible due to the critical nature of the application, especially when the vendor providing the application guarantees correct operation only in a specific configuration. In this paper, we propose a method to solve this problem. The idea is to run unpatched and patched application instances concurrently, with the unpatched one having complete control and the output of the patched one being used only for comparison, to watch for differences that are consequences of introduced bugs. To test this idea, we developed a system that allows us to run web applications in parallel and tested three web applications. The experiments have shown that the idea is promising for web applications from the technical side. Furthermore, we discuss the potential limitations of this system and the idea in general, how long two instances should run in order to be able to claim with some probability that the patched version has not introduced any new bugs, other potential use cases of the proposed system where two application instances run concurrently, and finally the potential uses of this system with different types of applications, such as SCADA systems.
\end{abstract}


\section{Introduction}
\label{sec:introduction}

Software patching is an increasingly important aspect of keeping today's computing environment secure, where the volume, complexity, and number of configurations under which software runs have increased significantly \cite{dadzie2005}. However, it is not always possible to patch all systems for a variety of reasons. A common reason is that the system in question is critical and/or is only certified for a particular configuration. Among other reasons, the software vendor may prohibit changes to the software for fear that they could cause harm. This is especially the case for mission-critical applications, such as control systems, where failures in system execution can lead to major losses and even loss of life. In particular, patching of ICS devices is usually postponed to scheduled production outages to prevent potential operational disruption of critical systems \cite{chen2018}.

In this paper, we present an idea on how to address this challenge, i.e., how to patch systems in a safe way that can be applied to any kind of mission-critical systems. The core idea is to run two systems in parallel, with the only difference being that one is patched and the other is not. The unpatched system is used normally, while the patched system receives the same input as the unpatched system and its output is compared to that of the unpatched system. If there is a difference between the two outputs, then the patch may have introduced an error. If there is no difference between the outputs within a certain period of time, it can be assumed with a certain probability that patching has not caused any changes in the behavior of the system and hence it can be put into production.

We present a system we developed to test this idea on web applications. It is a simplified system, but it shows the basic concepts and allows us to do some experiments and develop the idea further. To be able to send a single request to two identical components and then compare their outputs, we built a proxy that sits in front of the patched and unpatched components and controls the communication between them and the outside world while watching for possible differences in the response.
%ID > use caseovi ukratko 

Besides secure patching, there are several other use cases for the system developed in this work. Due to its parallelization capabilities, it can be used in differential testing. In addition, it can be used to detect attacks, since the changes in the output do not necessarily have to be related to a bug. An attacked system that has been compromised could produce different output for the same inputs. We discuss these potential uses further in Section \ref{sec:discussion}.

The paper is structured as follows. In Section \ref{sec:systemarchitecture} we describe the architecture and implementation of the system we used to perform the experiments. In Section \ref{sec:experiments} we describe experiments we did and the lessons learned. In Section \ref{sec:discussion} we analyze the results and discuss what it would be necessary to do in other to support other types of systems, like components of industrial control systems. We also describe other use cases for such a system besides testing patched systems. In Section \ref{sec:relatedwork} we review related work. The paper finishes with conclusions in Section \ref{sec:conclusion} and list of references.

\section{System Architecture and Implementation}
\label{sec:systemarchitecture}

In this section we describe the idea on which the system is built, the architecture of the system we built, and some important details of the implementation. The results of the experiment are described in the following section.

The core idea for solving the problem of detecting behavioral changes in the system introduced by an update is to run both the unpatched and patched applications in parallel for a period of time and watch for unexpected behavioral deviations. The longer the applications run in parallel and the more different clients access them, the higher the probability that the update did not change the behavior of the software if no differences were found. Note that some differences are to be expected and should be accounted for, such as random values like tokens and cookies for web applications that use them, but otherwise there should be no differences.

\begin{figure}
    \centering
    \includegraphics[width=8cm]{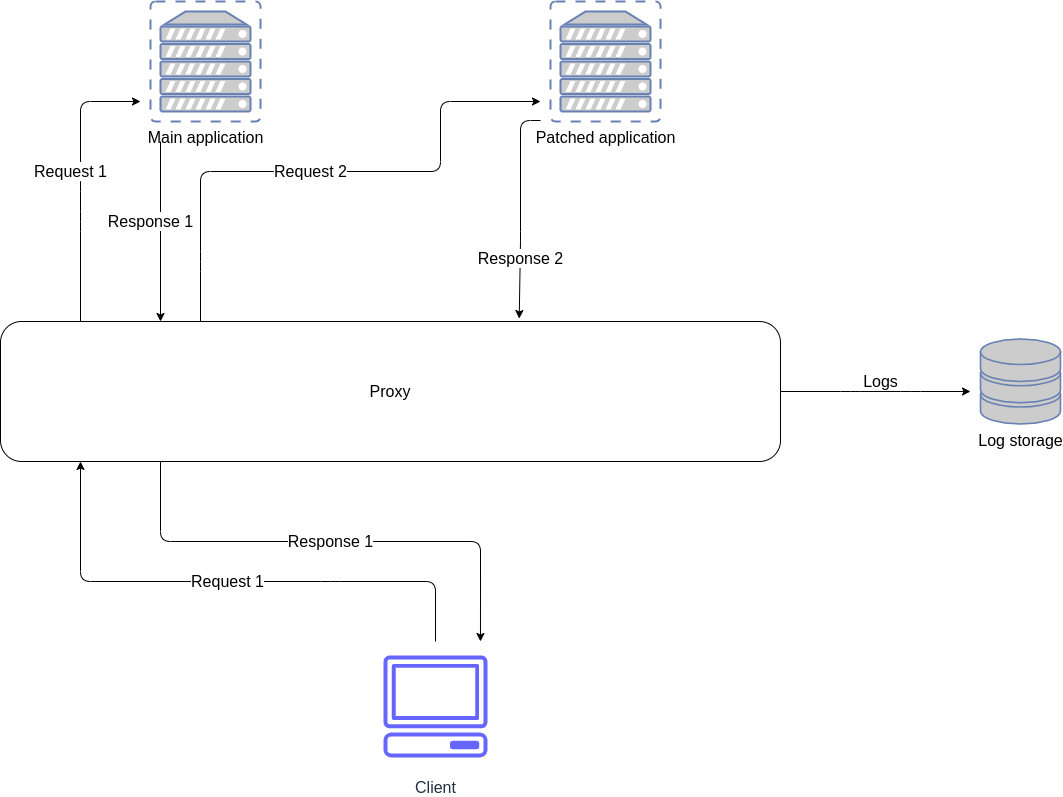}
    \caption{System Architecture}
    \label{fig:system_architecture}
\end{figure}

\begin{figure}
    \centering
    \includegraphics[width=8cm]{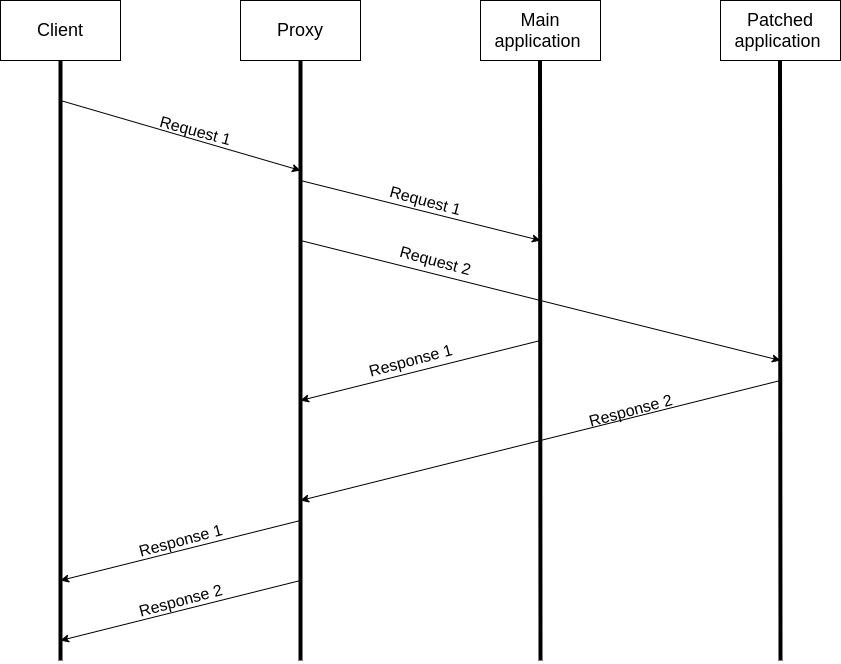}
    \caption{Communication sequence diagram for system shown on Figure \ref{fig:system_architecture}}
    \label{fig:communication_flow}
\end{figure}

The system architecture is shown in Figure \ref{fig:system_architecture}, while Figure \ref{fig:communication_flow} shows the communication flow in the system. The key component is
the \textit{Proxy} that intercepts traffic between the Client and two software instances. One software instance is the \textit{Main application} that is known to work without known bugs (except, possibly, the ones corrected by the patch) and is not patched, while the other software instance is the \textit{Patched application} for which we want to confirm that it is working correctly, too. The Proxy sends requests from the Client to the Main application (\textit{Request 1}) and sends adjusted copies of those requests to the Patched application (\textit{Request 2}). Then, when it receives responses from both application components (\textit{Response 1} and \textit{Response 2}), it passes the response from the Main application back to the Client. At the same time, it compares responses from the main application and the patched application, ignoring expected differences. If there is an unexpected difference, it signals an alarm, because a difference in behavior between the two has been found.

% Mitmproxy is used for the proxy implementation. 

The proxy can be used in what is called \textit{learning mode}. The goal of this mode is to find the expected differences that are ignored during testing. In this mode, two identical, i.e. unpatched, application instances that are known to work without known errors, are run simultaneously, and the proxy detects and logs differences. The detected differences can be automatically or manually converted into rules for the proxy component that specify where the differences between two applications are expected to occur.

\begin{figure}
    \centering
    \includegraphics[width=8cm]{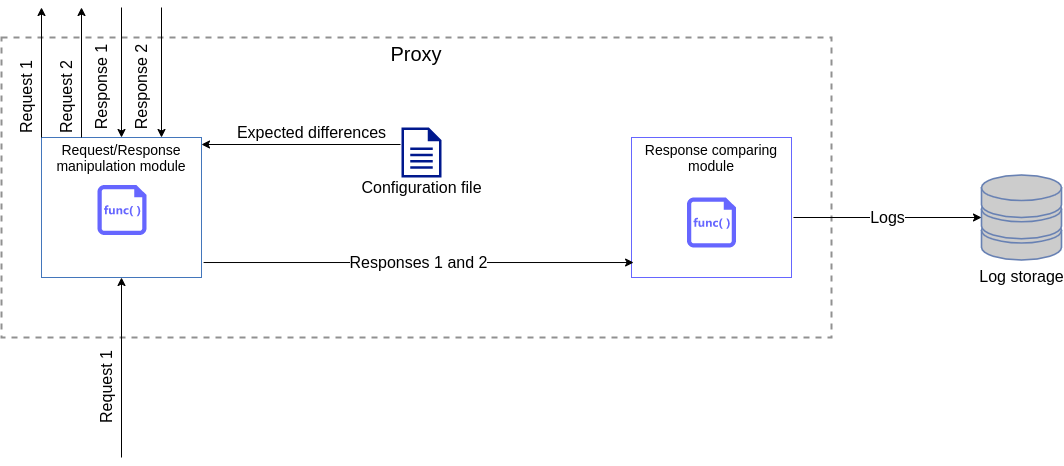}
    \caption{Proxy architecture}
    \label{fig:proxy_arch}
\end{figure}

The architecture of the proxy component is shown in Figure \ref{fig:proxy_arch}. Proxy consists of two modules: (\textit{i}) \textit{Request/Response Manipulation Module}, and (\textit{ii}) \textit{Response Comparing Module}. Mitmproxy \cite{mitmproxy} was used to implement proxy so both modules are implemented as plugins within the Mitmproxy.

% Mitmproxy is used for the proxy implementation. 

The Request/Response Manipulation Module (\textit{RRMM}) is a Mitmproxy script that intercepts traffic between the user and the originating application. The RRMM communicates with the response comparison module (\textit{RCM}) by forwarding the intercepted responses. Any request that targets the original application is replicated and customized for the patched application container. The customizations can consist of manipulations of both the request header and the request body. The manipulations consist of mapping feature values from the original application to the values used in the patched application. The user specifies the characteristic values in the configuration file. Characteristic values are a subset of expected differences that represent random pieces of data specific to each instance of an application, such as tokens and cookies. Response scraping is built into the module to retrieve the assigned values. The order of the responses does not have to match the order of the requests. To ensure that the RCM receives the correct response pairs, the responses are mapped by certain keys. The keys are inserted into the request headers, enabling the system to associate the original request with the replicated request. A correct response pair represents two responses for such original and replicated requests, the first targeting the original application and the second targeting the patched application. Each response pair is forwarded to the RCM.

The forwarding of the response pairs is not done in real time. The reason for this is that the responses in a pair do not have the same response time, with one of them possibly having a significant delay that could block the comparison system. For this reason, the user sets the comparing rate. For each time interval specified in the comparing rate, the RRMM forwards response pairs that have arrived up to that time to the RCM, which compares all received response pairs.

The RCM compares the responses and logs differences. The comparison of pairs of responses is performed at the specified comparing rate. Differences are expected if they are part of randomly generated data specific to each client session, such as cookies and tokens. The RCM can be divided into two working modes, learning mode and comparison mode. 

The learning mode is based on the formatted RCM output files. The RCM outputs the differences and the context of the response in which the differences occur, such as the html tag in which the token is placed. From the output of the differences, the user can learn about the differences and identify expected differences. Expected differences can be detected from keywords and differences in numbers that depend on randomization, while unexpected differences can indicate errors and vulnerabilities. In this case, the comparing module generates an alarm. From the alarm, one can extract the necessary information about the differences and check whether there is a bug in the code or a vulnerability.

Comparing mode is the inner process of the RCM that finds differences. Various methods are used for response comparison. Responses such as HTML and JavaScript can be compared as plain text. For plain text comparison, Diff Match and Patch \cite{diff-match-patch} is used. Diff Match and patch is a Google library packaged for Python. Differences are stored as tokens in diff files.

The RCM finds differences from stored tokens. The token represents the response part that is different in the application instances. For each difference, there are two tokens, one for each instance. The RCM then searches the response to find the location of the token. If the difference token is associated with one of the legal differences, for example in the same tag whose attribute is included in the set of expected differences, then the difference is expected.

JSON responses are compared as objects based on the values of their attributes, since JSON is a text-based, human-readable format that describes an object. With JSON, the same object can be described in multiple ways, with differences in, for example, the order of attributes. Moreover, an object can contain objects that in turn contain other objects, and so on. For this reason, the RCM iterates recursively through the object hierarchy when making a comparison. This allows the RCM to perform a deep-compare, that is, it can compare differences even at the lowest level of the object attribute tree.

In order for the RCM to know which differences are expected and which are unexpected, descriptions of the expected differences must be produced. To do this, the RCM provides a special output. From the special output, the user can extract the expected differences. An example of a special output is shown in Listing \ref{lst:special-output}.

\begin{lstlisting}[caption=Specialized output for extracting differences,label=lst:special-output]
token: b0008cae11221c91b9a8f65d17c33202c819c5fa found in tag:
<script type="text/javascript">
  var odoo = {
    csrf_token: "b0008cae11221c91b9a8f65d17c33202c819c5fao",
        };
</script>
\end{lstlisting}

The expected differences and the characteristic values are specified in the configuration file. The characteristic values can be extracted by examining the traffic, especially the request bodies. Web applications can send characteristic values, such as CSRF tokens, in request bodies. CSRF token is a secret, randomly generated value that the server-side applications generate for the client. When the client later makes a request, the token is validated. CSRF tokens are used to protect against CSRF attacks.

The configuration file is simple, as shown below. It consists of legal differences and characteristic values. Legal differences are indicated by a ':' character at the beginning of the line. Characteristic values are marked with a '+' sign at the beginning of the line.

\begin{lstlisting}[caption={Configuration file},captionpos=b]
:csrf_token
:session_info
:id
:last_update
:search_view
:date
:context
+csrf token
\end{lstlisting}

\section{Experiments}
\label{sec:experiments}

Experiments were conducted on three applications: eShopOnWeb \cite{eshoponweb}, Odoo \cite{odoo} and OpenProject \cite{openproject}. The goal of the experiments was to test the idea over real systems, and to detect any details that we were not aware of, such as some crucial show stoppers. We also wanted to see if the number of details that we need to consider is constant for each new application, or solved after a small number of applications tested.

The first step towards parallelization is to run two instances of the unmodified application. Docker containers \cite{docker} were chosen for this purpose. A container is a standard software unit that packages the code and all its dependencies in such a way that the application can be quickly and reliably transferred from one computing environment to another. Running two separate Docker containers is equivalent to running two instances of an application.

\subsection{eShopOnWeb}

EShopOnWeb is a sample ASP.NET Core web shop application powered by Microsoft. The main purpose of the application is to demonstrate patterns and principles of web development \cite{eshoponweb}. 

The application represents a simple web shop application. Its set of functionalities consists of user login, adding items to cart, purchasing items, editing user profile, etc. HTTP traffic of the application consists of GET and POST requests and responses with HTML body form. JSON traffic and other types of body forms do not occur. Although the set of functionalities is limited, it has proven to be a huge source of information. 

The first part of the response that we identified as an expected difference is the Request Verification Token. For security reasons, the application generates request verification tokens in GET and POST responses, which the client must return in the next GET or POST request. Request Verification Tokens are random values and each application instance generates a different value in each request. Therefore, patched and unpatched applications return different values for request verification tokens. This means that we need to extract the token's value from the patched application, and when the client makes the next request for the unpatched application, we need to replace the token's value in the request copied for the patched application with its characteristic token that we remembered earlier. Cookies are also specific to each application. Each instance of the application generates its own cookies. Cookies are sent via headers, specifically \textit{Set-Cookie} header field. 

POST requests usually have content in the request body or in the request query. When encoding the content, \textit{percent encoding} \cite{percent-encoding} is used.

Although the complexity of application traffic is mostly low, there is no JSON serialized data and no multipart form data. Nevertheless, it is complex enough to recognize the patterns for application parallelization. The proxy configuration file for eShopOnWeb is shown in Listing \ref{lst:lst_3}. In addition to the RequestVerificationToken mentioned earlier, another legal difference is the item identifier. EShopOnWeb is a web shop application based on actions with items. Items represent assets that can be purchased in the shop. Different instances of the application use different databases. Consequently, the same items can have different identifiers in the database. When an item is added to the cart, the item identifier in the request body is forwarded and a different identifier is forwarded for the other instance.

\begin{lstlisting}[numbers=none, caption={eShopOnWeb configuration file}, label={lst:lst_3}, captionpos=b]
:RequestVerificationToken
:Items[
+RequestVerificationToken
+Items[
\end{lstlisting}

\subsection{Odoo}

Odoo is a business management software that includes CRM, e-commerce, billing, accounting, manufacturing, warehouse, project management, and inventory management \cite{odoo}.

Odoo is open source and has Docker support. It is more complex and has a greater number of functionalities than eShopOnWeb. The traffic is different from that generated in eShopOnWeb. Pure HTML responses are in the minority. JSON is most commonly used \textit{Content-Type} in requests and responses. 

JSON objects are easy to read and copy, so parallelization is not a problem. The proxy simply copies the content and forwards it to the application containers. The problem occurs when comparing responses. The responses can contain a lot of information, sometimes a list of JSON objects that have other JSON objects as attributes. In this case, the order of the attributes may not be the same in both responses. When the responses are compared as text using a simple diff algorithm, a difference is detected and a false positive alarm is raised. For this reason, JSON responses are compared as objects, on an individual attribute basis. This way, only real differences (e.g. date, ID, update, etc.) are detected and logged.

Another expected difference that occurs in eShopOnWeb application traffic relates to HTML JSON attributes. It is possible for two responses to have the same HTML code, but be marked as different due to a different order of attributes. We solved this problem by extracting all the characters from the tags that are different in the responses and storing them in two character lists. Then we excluded the non-alphabetic characters from the character lists and sorted them in alphabetical order. If the sorted character lists are the same, the responses are treated as equal. This approach assumes that there are no permutations of parameter names, since different attributes such as "name" and "eman" will give the same result when sorted in alphabetical order. The probability of this being the case is negligible.

% eksplicitno opisati sortiranje i postupak uspoređivanja, pretpostavka da neće biti permutacija od istih riječi, koji će dati važeči html drugog značenja
% skratit bez detalja o implementaciji

The Odoo proxy configuration file is shown in listing \ref{lst:lst_4}. \textit{Session\_info} contains data about the user session. Each user session has specific values are different for each user session. \textit{Write\_date}, \textit{create\_date}, \textit{date}, and \textit{last\_update} are time-dependent parameters. Requests sent by the client to the unpatched version and the request created by the proxy sent to the patched version do not arrive at their destination at the same time. For this reason, the time-dependent parameters are different for different instances of the application running in parallel. The Id parameter represents a randomly generated string that varies from instance to instance. The Id parameter appears in html tags as an attribute.

When all of these factors are incorporated into the proxy, parallelization is achieved. Moreover, the logs of differences show that only legal differences are displayed when we use two instances of the same version of the application. This is the expected and desirable result.

\begin{lstlisting}[numbers=none, caption= {Odoo configuration file}, label={lst:lst_4}, captionpos=b]
: csrf_token
: session_info
: write_date
: create_date
: id
: last_update
: search_view
: date
+ csrf_token

\end{lstlisting}

\subsection{OpenProject}

OpenProject is a project management software. Functionalities include: project planning and scheduling, product roadmap, release planning, task management, release collaboration, Agile, Scrum, time tracking, cost reporting, budgeting, bug tracking, wikis, forums, meeting agendas and meeting minutes. OpenProject is the leading open source project management software \cite{openproject}. 

The Docker image is publicly available, so setting up two running instances of OpenProject in parallel is not a problem. The web applications mentioned above, eShopOnWeb and Odoo, did not use multipart form as a content type, so this application comes with a new implementation challenge. Multipart form is a content type where the request body is specifically formatted as a series of "parts" separated by MIME (Multipurpose Internet Mail Extensions) boundaries. The multipart form allows one or more different sets of data to be combined into a single body, such as a text file for upload and string parameters related to the uploaded file.

To achieve parallelization of multipart requests, a multipart form encoder and decoder implementation is required. Open source implementations of this type are used for this purpose. The goal of parallelizing multipart form requests is successfully achieved.

The OpenProject proxy configuration file is shown in Listing \ref{lst:lst_5}. OpenProject uses the authencity token, which is the Ruby alternative for the CSRF token \cite{chaudhuri2010}. Its purpose is the same. CreatedAt and updatedAt are timestamp parameters that differ for application instances because of the latency introduced by the proxy. When the proxy intercepts a request, it takes time to replicate and adjust the new request. The original application and the patched application do not receive their requests at exactly the same time. Styles is a parameter that specifies the static CSS files to be used in the application. CSS file names sometimes contain a randomly generated string that is different for each application instance. Nonce is an arbitrary number that can be used only once in the communication \cite{rogaway2004}. For each application instance, nonce is different because it is an arbitrary number, so it is an expected difference.

\begin{lstlisting}[numbers=none,caption={OpenProject configuration file},label={lst:lst_5},captionpos=b]
:token
:nonce
:uuid
:key
:styles
:timestamp
:createdAt
:updatedAt
+token
+nonce
\end{lstlisting}

\section{Discussion}
\label{sec:discussion}

%topics
% kao kraj uvoda, poglavlje je podijeljeno na potpoglavlja, o svakoj uvodna riječ
In this section we discuss several topics. First, in section \ref{subsec:reliability}, we discuss the issue of how long one should run the application instances in parallel to be able to assert with certain probability that the patched application instance is working correctly. In section \ref{subsec:issues} we discusses possible problems and restrictions of the proposed patching approach. Then, in section \ref{subsec:usecases} we discusses additional use cases of the system we prototyped besides patch management. Finally, since our future work related to the presented method is related to its application in industrial control systems, and specifically for updates of SCADA systems, we give a brief overview of our plans in section \ref{subsec:scadause}.

\subsection{Reliability}
\label{subsec:reliability}

%MATEJEVO POGLAVLJE

%We proposed a reliability model, but it is simple and it does not take into account the order of visited URLs. It also does not consider the use of different url get parameters to test the patched application. The model only evaluates url get parameters that are logged as visited url in the unpatched application.

For anyone who wants to use the method proposed in this paper, the question is how long patched and unpatched application instances should run concurrently before it can be claimed with some probability that their behavior is the same. This is a very important question, since running two instances for too short a period can leave undetected and potentially disastrous bugs. On the other hand, running them for too long wastes valuable resources, so this time should be minimized. This question is left for future work, but we will provide a sketch of a possible answer.

Suppose that when we started running the patched and unpatched applications concurrently, we found $N$ differences stemming from bugs in the patch, that we resolved. Further, suppose that the time between starting the concurrent execution and finding the $i$-th difference is $T_i$ time units, with $i$ being $i=1 \dots N$. To illustrate, when we first started the comparison of the patched and unpatched application instances, it took $T_1$ time units to find the first difference. After that, we took some time to analyze and correct the detected difference, after which we started concurrent execution again. Measured from this point in time, the next difference was detected after $T_2$ time units, and so on. Now suppose that after all $N$ differences have been found and resolved, we need to decide whether to stop concurrent execution or wait for the $N+1$-th difference to occur. To answer this question, we need to know the probability of executing two instances without finding any differences for a given time $T_{required}$. In other words, we are interested in finding out the probability that no differences are found in the interval $T_{required}$. For practical purposes, the person estimating the reliability of a system will set $T_{required}$ to some value, such as the time when the next patch is expected.

To calculate the probability of a difference NOT occurring in a given time we'll assume that the time between starting concurrent execution and difference occurring ($T_i$) follows an exponential distribution, $T\sim \mathcal{E}(\lambda)$ where $\lambda >0$, with probability  density  function (PDF) $f_T(t;\lambda)=\lambda e^{-\lambda\,t},\,t\geq0$. This assumption is taken from Driel et. al \cite{driel2020}. Furthermore, we equate differences in applications with bugs, i.e. we assume that differences result from bugs introduced by patch.

We estimate the parameter $\lambda$ from the times $T_i$ finding uniformly minimum-variance unbiased estimator (UMVUE), using method of moments estimator and point estimation theory (see \cite{bain2000}, Chapter 10.4, page 337), so that we get the estimator with minimum error. So, the UMVUE for $\lambda$ is
\begin{equation}
    \hat{\lambda} = \frac{N-1}{\sum_{i=1}^N T_i},
\end{equation}
with standard error (se) defined as standard deviation of $\hat{\lambda}$, i.e. 
\begin{equation}
    se(\hat{\lambda}) = \sqrt{\frac{\lambda^2}{N-2}}
\end{equation}

The question is what is the probability that no new differences occur in the given time $T_{required}$. More specifically, we are looking for an estimate of the time of occurrence of a new difference ($T$) after the detection has run for a given time. Mathematically, this event can be written as $\{T >T_{required}\}$, and we calculate the probability for this event. The calculation is given in the following expression:

\begin{equation} \label{eq1}
\begin{split}
P(T > T_{required}) &= 1 - P(T \leq T_{required})\\ 
   &=  1 - \int\limits_0^{T_{required}} \lambda e ^{-\lambda\,t}dt \\
   &= e ^{-\lambda\,T_{required}}
\end{split}
\end{equation}

So, to decide whether to stop with concurrent execution, the user of this system should do the following:

\begin{enumerate}
    \item decide on the value of $T_{required}$,
    \item calculate $\lambda$ and standard deviation based on $T_i$,
    \item using expression (\ref{eq1}) calculate probability,
    \item make a decision based on probability and standard deviation of $\lambda$.
\end{enumerate}

\subsection{Potential Issues \& Restrictions}
\label{subsec:issues}

The proposed method of patching certainly has some limitations and is definitely not applicable to every possible situation that might arise in the real world. Moreover, there are some other problems that we did not encounter in our experiments, but which are still highly likely to occur and need to be taken into account.

First, note that a patch could introduce a new behavior in the patched application that is different from the behavior of the unpatched application and fixes a bug. In this case, the proxy will detect a difference, but should not signal an error. A more interesting question is what to do in such a case. Should the proxy pass on the response of the unpatched application - which has an error - or that of the patched application, for which it is not yet certain whether it functions well? The answer to this question is probably context-dependent and should be decided by the users of the system, who know the patched application well.

The next problem might arise when the patch adds a new functionality. In this case, running the patched and unpatched instances simultaneously will not test the new functionality. In this case, of course, the new functionality needs to be thoroughly tested in the development and deployment phases, while the method presented in this paper will only help with functionality present in both application instances. In extreme cases, someone could patch the application to the point where its behavior is drastically different from the unpatched application. For example, the states that the application goes through might change, likely resulting in changes to the application's output. It seems reasonable to assume that the approach presented in this paper works best for patches that contain only bug fixes, and the more behavioral changes are introduced into the application, the less likely it is that this approach is feasible. This issue requires further and much deeper research.

The characteristics of the applications we used for the experiments must also be considered. First, they use the TCP protocol, which means that communication is reliable and messages are not reordered, duplicated or lost. Furthermore, while the applications implement complex state machines, they do so on top of a simple request-response protocol (i.e. HTTP) that runs in lockstep for both the patched and unpatched applications. If this were to change, there would be consequences to deal with:

\begin{itemize}
    \item using unreliable protocol, like UDP, would mean that the proxy would have to know that the messages could be lost, and in that case it should wait for retransmissions. Yet, the the difference could arise because the response wasn't sent at all, rather than the message being lost. This definitely adds some complexity to the implementation of the proxy.
    
    \item Using an application protocol that is more complex than HTTP would also increase the complexity of the proxy, to the point that the proxy would have to implement a part of, or even the entirety of the protocol, to be able to monitor for differences between the applications.
    
    \item A special case of the previous point would be a protocol that doesn't specify rigidly the order of messages, i.e. the one that allows messages to be interlaced. In that case, it might happen that the patched application sends message A, while the unpatched sends message B. Later, at some point, the patched application sends message B, while the unpatched sends message A. The two are equivalent, but handling such a situation is obviously not easy.
\end{itemize}

The conclusion is that the applicability of the proposed method depends heavily on the protocols used. Further research is needed to determine all the exact limitations and problems that may arise.

Finally, as we have seen in our experiments, randomness is a frequent source of differences. We were only dealing with one aspect of it, such as cookies, i.e. well-defined data whose definition allowed us to deal with them. But randomness can occur for a variety of reasons, including operating system scheduling, application dependence on sources other than users, inaccurate timers, deferred interrupts, etc. Dealing with them would also require further research on more complex applications than just the ones we use.

Due to all the issues discussed, changes in the application itself may be required to fully utilize this method. The goal of the changes would be to facilitate the creation of proxies that monitor updated applications. Further research is needed to understand what these changes are.

\subsection{Other potential use cases}
\label{subsec:usecases}

While we have intended this system primarily to enable patching of applications that are otherwise difficult to patch due to various restrictions, we have also envisioned two additional use cases. Both of these additional use cases expect a well-defined interface between an application and the outside world. For example, the interface could be the PSD2 OpenBanking API, or a firewall that blocks or passes packets through. 

The first use case of the proposed system is related to differential testing for the purpose of certification. Differential testing compares two different implementations of the same application logic to find bugs in the code. The idea here is that running two applications side-by-side would allow easier testing of the correctness of their implementation. This could simplify the certification process. Furthermore, if one application is a reference implementation of a functionality, other applications could be compared to it through differential testing.

The second use case is concerned with protecting against attacks on an application/system and other potential misuses of the application/system in question. For example, an attacker could compromise the firewall by sending specially crafted packets, or in the case of an API, send a request that could result in the application implementing the API being compromised. The idea is that applications and systems have bugs, but if they are implemented by different teams, there is little chance that they will have bugs in the same or similar parts of the application. Therefore, running two applications concurrently and comparing their outputs could be used to detect if an application/system has been compromised. Nonetheless, the causes of differences could be bugs rather than attacks. In any case, differences in responses of two applications that should behave the same way are a reason to investigate what happened.

\subsection{Application to other types of systems}
\label{subsec:scadause}

In our experiments, we only consider use cases involving web applications. However, there are many other types of systems, such as desktop applications, mobile apps, and industrial systems. These systems differ in their architecture and require different configurations of proxies and different types of response comparison. Simple desktop applications, for example, handle input and output through graphical user interfaces. 
In this section, a possible configuration is proposed for a specific family of systems, namely Supervisory Control And Data Acquisition (SCADA) systems.

Compared to IT systems, Industrial Control Systems (ICSs) are characterized by simpler and more deterministic network traffic containing smaller well-ordered packets \cite{galloway2012}. This should, at least in theory, make both the space of potential inputs and the state space smaller than in IT systems, leading to potentially better performance of our approach on ICSs.

To apply the method presented in this paper to patching of SCADA systems, some additions are needed. As before, patched and unpatched systems - in this case SCADAs - should run concurrently. The unpatched version produces outputs that control physical processes, while equivalent output of the patched version is monitored. We assume that the unpatched and patched SCADA systems run on separate, otherwise completely identical machines. Also, each of them has a human-machine interface (HMI). In this case, two things need to be compared, first, that the HMIs produce the same output, and second, that the same control commands are sent to the same devices on the communication channel used for data acquisition and control.

\begin{figure}
    \centering
    \includegraphics[width=8cm]{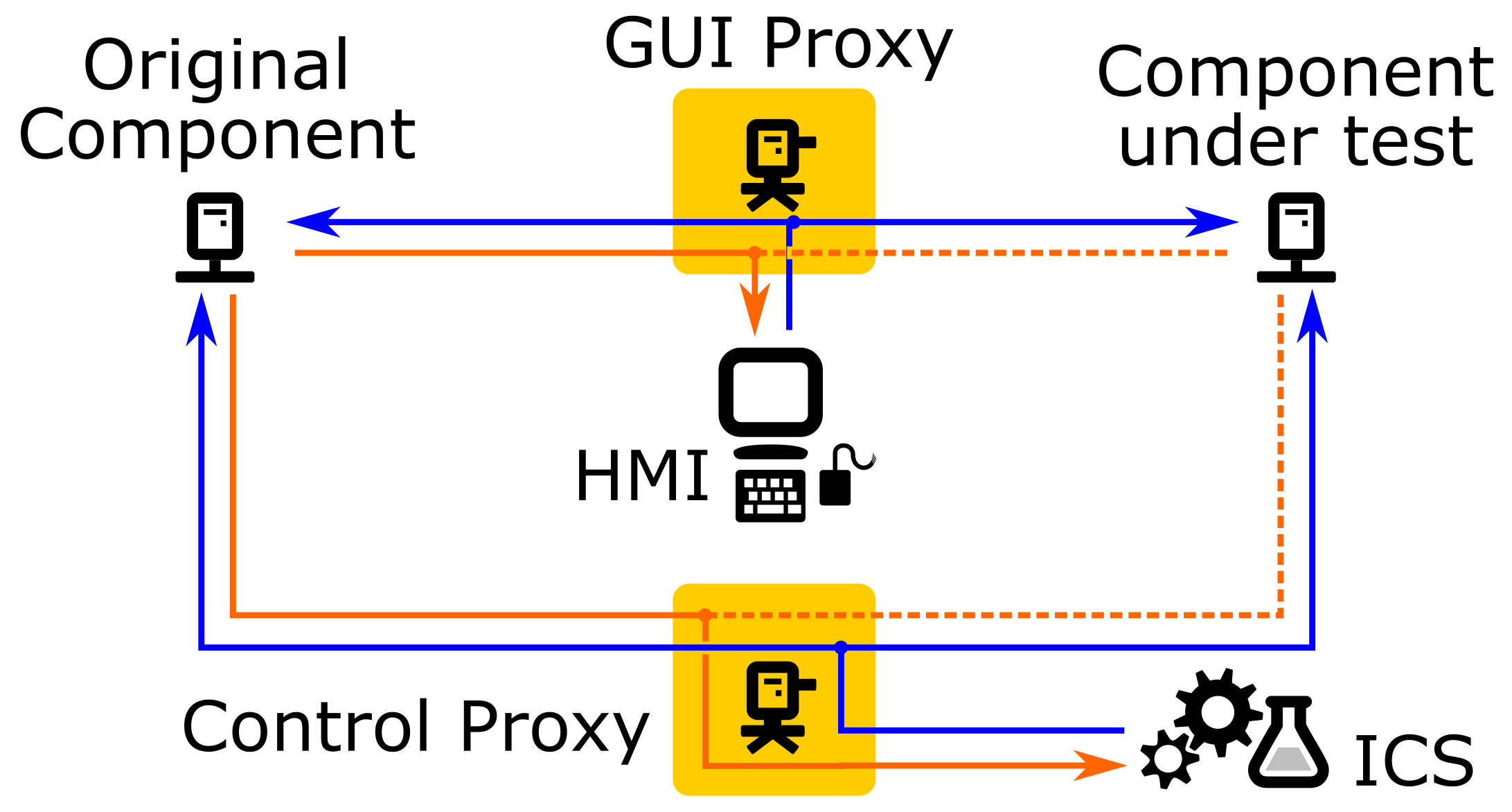}
    \caption{Study of the use case with SCADA systems}
    \label{fig:scada_example}
\end{figure}

To illustrate this system, we'll use Figure \ref{fig:scada_example}. There are two communication channels that each SCADA system uses. One communication channel exists between the SCADA systems and the industrial control system (ICS) that contains other components and PLCs. The other communication channel is between the human user and the SCADA system. Through this communication channel, the user receives information about the current state of the system and gives commands on how to change the state of the system. Since there are now two communication channels, this means that two proxies are required. The \textit{control proxy} compares control requests from the unpatched and patched SCADA systems. The request from the unpatched version is propagated to the ICS, while the request from the patched version is used only to detect differences. In Figure\ref{fig:scada_example}, the requests and data sent to the patched and unpatched systems are represented by blue lines, while the responses and control signals from these systems are represented by orange lines. Dotted lines represent responses that are only used only by the proxies and are not propagated to their final destinations.

The other communication channel is towards the users, through which they control the physical system via the human-machine interface. The \textit{graphical user interface proxy} (GUI proxy) records the user's actions, replicates them to the patched instance of the SCADA system, and compares the information returned by these systems to the user. This includes cursor movements, mouse clicks, and any actions on the front-end component. It should be noted that the information being compared will also be graphical, in which case the GUI proxy should be able to properly analyze graphical data. The user is presented with the response of the unpatched version, while the response of the patched version is compared with the unpatched version to detect potential differences.

% Mitmproxy is used for the proxy implementation. 

\section{Related Work}
\label{sec:relatedwork}

When searching for related work, we identified several similar systems, but none of them was identical to ours, or had the same purpose. Some related areas include:
\begin{enumerate}
    \item differential testing,
    \item asynchronous patching,
    \item differential static analysis and symbolic execution,
    \item differential fuzzing and hybrid techniques,
    \item and regression testing.
\end{enumerate}

The idea of testing software by comparing it to functionally equivalent codes was first proposed by McKeeman \cite{mckeeman1998} in 1998. This method is a form of random testing called differential testing. Differential testing requires that two or more comparable systems are available to the tester. These systems are presented with an exhaustive set of mechanically generated test cases. If the results differ or one of the systems goes through an infinite loop or crashes, the tester has a candidate for a bug-exposing test \cite{mckeeman1998}.

There is a connection between our method and differential testing. The idea of testing comparable systems in differential testing can be compared to testing a patched and an unpatched application. Patching software creates two application instances that may differ in certain areas, which makes them comparable systems. The main difference between our approach and differential testing is that the comparable systems mentioned in \cite{mckeeman1998} are different implementations that serve the same purpose, while in this paper we propose the comparison of different versions of the same implementation. Moreover, while the motivation behind differential testing is to find bugs in different implementations of the same idea, the motivation of this paper is to study the consequences of application updates by running different versions of the same application in parallel.

% Parallel (async) patching
The concept of patching parallel software instances is not new. For example, the system described in \cite{mcneill2014} and \cite{mcneill2016} performs patching over a copy of a target virtual machine (VM) in a special maintenance environment, while another copy of that VM is accessible to operational clients. After the patches are successfully applied, the patched copy of the VM is made available to clients. However, unlike our method and system, their system does not analyze the differences between the patched and unpatched versions of the VM and does not automatically verify that the patches have not affected important functionality.

% Differential static analysis & symbolic execution
Approaches such as \cite{srivastava2011} and \cite{palikareva2016} use static analysis and/or symbolic execution to determine the differences between multiple software instances. These methods require access to the source codes or binaries of the instances and often find differences that do not affect the functionality of the software. In contrast to these approaches, we do not require access to the code of the instances and are not interested in differences that do not affect the functionality of the software in question.

% Differential fuzzing & hybrid techniques
Other approaches use differential fuzzing \cite{walz2017,argyros2016} and combinations of fuzzing with symbolic execution \cite{brumley2007,cadar2008,petsios2017,noller2020,jin2010} to find test cases for which software instances that are supposed to implement the same specification produce different results. These test cases are often targeted towards simple communication sequences, such as establishing a TLS session, and aim to find edge cases that affect security. In contrast to this group of approaches, we focus on testing software instances over longer periods of time, using realistic user sessions and load rather than generated tests, to determine if a patch has introduced new bugs that affect practical functionality. Unlike \cite{brumley2007,cadar2008,noller2020,jin2010}, we do not require access to the code of the software instances.

% Regression testing
Many of the approaches mentioned earlier, such as \cite{palikareva2016,noller2020,jin2010}, support test generation for regression testing. In regression testing, the same collection of tests is run on the old and new versions of a software to detect newly introduced bugs. We are similar to some extent to regression testing, but the main difference between regression testing and our approach is that we test with real requests rather than predefined or generated tests.

\section{Conclusions}
\label{sec:conclusion}

In this paper, we have proposed a system for controlling updates of software components. We have designed a system that can monitor two web application instances, one representing the unpatched version of the application and the other representing the patched version of the application.

With this approach, the user can extract the expected differences in the behavior of two application instances. If the applications have logs of visited URLs, one can find the differences between the outputs of the two instances and evaluate the reliability of the patched version of the application. The system was tested on three web applications, with the expected differences determined experimentally.

The main challenge in developing the system was to enable meaningful parallel use of the application instances. Different web applications involve different forms of communication, such as different content types and encodings, so it is a major challenge to implement a general parallelization system.

As future work, the system can be integrated into an automated testing framework so that the reliability assessment process can be faster. Also, the system could allow the user to change the version of the application they are interacting with. In this work, we always interacted with the unpatched version of the application through the web client, but it could be useful to change the version of the application in the user session.

\IEEEtriggeratref{19}

\bibliographystyle{IEEEtran}
\bibliography{bibliography}

\end{document}